\documentclass[pra, twocolumn,preprintnumbers,amsmath,amssymb,showpacs,aps,pra]{revtex4}
\usepackage{mathrsfs}

\usepackage{graphicx}
\usepackage{dcolumn}
\usepackage{amsmath}
\usepackage{epsfig}
\usepackage{subfigure}
\usepackage{color}
\usepackage{epstopdf}
\usepackage{framed}

\begin{document}

\date{\today}

\title{Entanglement-Enhanced Sensing in a Lossy and Noisy Environment}%
\author{Zheshen Zhang}%
\email{zszhang@mit.edu}%
\author{Sara Mouradian}
\author{Franco N. C. Wong}
\author{Jeffrey H. Shapiro}

\affiliation{Research Laboratory of Electronics, Massachusetts Institute of Technology,
77 Massachusetts Avenue, Cambridge, Massachusetts 02139, USA}%

\begin{abstract} 
Nonclassical states are essential for optics-based quantum information processing, but their fragility limits their utility for practical scenarios in which loss and noise inevitably degrade, if not destroy, nonclassicality.  Exploiting nonclassical states in quantum metrology yields sensitivity advantages over all classical schemes delivering the same energy per measurement interval to the sample being probed.  These enhancements, almost without exception, are severely diminished by quantum decoherence. Here, we experimentally demonstrate an entanglement-enhanced sensing system that is resilient to quantum decoherence. We employ entanglement to realize a 20\% signal-to-noise ratio improvement over the optimum classical scheme in an entanglement-breaking environment plagued by 14\,dB of loss and a noise background 75\,dB stronger than the returned probe light.  Our result suggests that advantageous quantum-sensing technology could be developed for practical situations. 

\end{abstract}

\maketitle

Quantum information processing (QIP) exploits fundamental quantum-mechanical properties to realize capabilities beyond the reach of classical physics. Nonclassical states are essential for optics-based QIP, providing the bases for quantum teleportation \cite{bennett93, bouwmeester97, ma12}, device-independent quantum key distribution \cite{acin07}, quantum computing \cite{shor97,vandersypen01}, and quantum metrology \cite{giovannetti11}. Nonclassical states can increase the signal-to-noise ratios (SNRs) of quantum-metrology systems. Indeed, squeezed states have been employed to beat the classical-state limits in optical-phase tracking \cite{yonezawa12,humphreys13}, biological sensing \cite{taylor13}, and gravitational wave detection \cite{aasi11,aasi13}. Squeezed states, however, are vulnerable to loss:  a 10\,dB SNR enhancement without loss degrades to 1\,dB in a system with 6\,dB of loss.  Under ideal conditions, N00N states yield SNR improvements comparable to those of squeezed states \cite{dowling98,afek10,xiang10,ono13}, but noise injection can easily render N00N states impotent in this regard \cite{huelga97,escher11}.  Consequently, quantum decoherence, arising from environmental loss and noise, largely prevents any quantum-sensing performance advantage, casting doubt on the utility of QIP systems for practical situations.

Quantum illumination (QI) is a radically different paradigm that utilizes nonclassical states to achieve an appreciable performance enhancement in the presence of quantum decoherence. QI can defeat eavesdropping on a communication link \cite{shapiro09,xu11,zhang13,shapiro14}, and boost the SNR of a sensing system \cite{lloyd08,tan08,shapiro-lloyd09,s.zhang14,barzanjeh14,guha09,lopaeva13}. QI systems are comprised of: (1) a source that emits entangled signal and idler beams; (2) an interaction in which the signal beam (used as a probe) is subjected to environmental loss, modulation, and noise en route from the source to the receiver; and (3) a receiver that makes a joint measurement on the returned signal beam and the idler beam, which has been stored in a quantum memory, to extract information about the environment's modulation of the signal.  QI's performance advantage over classical schemes of equal probe energy derives from the fact that QI's initial entanglement---although destroyed by the lossy, noisy environment---creates a correlation between the returned and retained light that is much stronger than what can be obtained with classical resources.  The joint-measurement receiver relies on this stronger signature to achieve its better-than-classical performance. The QI-enabled performance advantage in a secure communication system was demonstrated in \cite{zhang13}: the measured bit-error rate (BER) for the legitimate parties in that experiment was five-orders-of-magnitude lower than the BER suffered by the passive eavesdropper. 

Demonstrating QI's performance advantage in sensing is a nontrivial task. The dramatic BER disparity that has been demonstrated in QI-based secure communication results from the legitimate users having access to the initial entangled state, while the eavesdropper can only measure weak thermal states.  In optimal classical sensing, however, there should be no restrictions on the transmitter other than its output energy, and there should be no restrictions on the receiver structure. Consider the QI target-detection experiment reported by Lopaeva \emph{et al}.~\cite{lopaeva13}.  It exploited the signal-idler photon pairs produced by spontaneous parametric down-conversion (SPDC) and a coincidence-counting receiver to demonstrate an SNR improvement over probing with a weak thermal state. But a coincidence-counting receiver does not fully exploit the entanglement of SPDC's signal and idler outputs.  Moreover, neither a weak thermal-state probe nor a coincidence-counting receiver represent optimum choices for classical-illumination (CI) target detection, which are known to be a coherent-state probe and a homodyne-detection receiver \cite{tan08,guha09}. Hence, experimental evidence for QI's target-detection advantage over an optimum CI system has yet to appear. In this paper, we provide that missing evidence by showing that QI can yield an appreciable target-detection performance gain---over an optimum CI system of the same transmitted energy---in an entanglement-breaking environment plagued by 14\,dB of loss and a noise background 75\,dB stronger than the returned probe light. 

In QI target detection using the entangled Gaussian states produced by continuous-wave (cw) SPDC \cite{tan08,shapiro-lloyd09,guha09}, the broadband signal (S) and idler (I) beams can be taken to be a collection of $M = TW$ independent, identically distributed signal-idler mode pairs, where $T$ is the duration of the measurement interval and $W$ is the phase-matching bandwidth. Each mode pair (with annihilation operators $\hat{a}_{\rm S}$ and $\hat{a}_{\rm I}$) is a two-mode squeezed state (TMSS) with Fock basis representation 
\begin{equation}
|\psi\rangle_{\rm SI} = \sum_{n = 0}^\infty \sqrt{\frac{N^n_{\rm S}}{(N_{\rm S}+1)^{n+1}}}|n\rangle_{\rm S}|n\rangle_{\rm I},
\end{equation} 
where $N_{\rm S}$ is the source brightness (mean photon number per mode). Signal-idler entanglement is then quantified by the phase-sensitive cross correlation (PSCC) of the TMSS, $ \langle\hat{a}_{\rm S} \hat{a}_{\rm I}\rangle = \sqrt{N_{\rm S}(N_{\rm S}+1)}$.  This PSCC equals the quantum limit for a mode-pair with $\langle \hat{a}_{\rm S}^\dagger\hat{a}_{\rm S}\rangle = \langle \hat{a}_{\rm I}^\dagger\hat{a}_{\rm I}\rangle = N_{\rm S}$.  For the usual $N_{\rm S} \ll 1$  cw-SPDC operating regime, it greatly exceeds the classical-state PSCC limit, $N_{\rm S}$, under the same average photon-number constraints.  

In our experiment, we first phase-modulate the signal modes.  Then we probe a weakly-reflecting target---which is embedded in a strong thermal-state background---with these phase-modulated modes, while the idler modes are stored in a quantum memory.  At the input to our joint-measurement receiver we then have available, for each SPDC mode pair, the signal-return mode, $\hat{a}_{\rm S}^{\rm in}(\varphi) =\sqrt{\kappa_{\rm S}}\,e^{i\varphi}\hat{a}_{\rm S} + \sqrt{1-\kappa_{\rm S}}\,\hat{a}_{\rm B}$, and the stored-idler mode, $\hat{a}_{\rm I}^{\rm in} = \sqrt{\kappa_{\rm I}}\,\hat{a}_{\rm I} + \sqrt{1-\kappa_{\rm I}}\,\hat{a}_v$. Here: $\varphi$ is the signal-mode phase shift; $\kappa_{\rm S}$ and $\kappa_{\rm I}$ are the roundtrip-probe and idler-storage transmissivities; $\hat{a}_{\rm B}$ (with $(1-\kappa_{\rm S})\langle \hat{a}_{\rm B}^\dagger\hat{a}_{\rm B}\rangle = N_{\rm B} \gg 1$) is the background mode; and $\hat{a}_v$ is the vacuum-state mode associated with idler-storage loss.  The joint-measurement receiver uses a low-gain ($G -1\ll  1$) optical parametric amplifier (OPA) to obtain the idler-mode output $\hat{a}_{\rm I}^{\rm out}(\varphi) = \sqrt{G}\,\hat{a}_{\rm I}^{\rm in } + \sqrt{G-1}\,\hat{a}_{\rm S}^{{\rm in}\dagger}(\varphi)$ for each mode pair at its input.  Direct detection of all $M$ idler-mode outputs from the OPA then yields the QI measurement $\hat{N}_{\rm I}^{\rm out}(\varphi)$, whose signal-to-noise ratio, 
\begin{equation}
{\rm SNR}_{\rm QI} \equiv \frac{4\langle \hat{N}_{\rm I}^{\rm out}(0) - \hat{N}_{\rm I}^{\rm out}(\pi)\rangle^2}{\left(\sqrt{{\rm Var}[\hat{N}_{\rm I}^{\rm out}(0)]} + \sqrt{{\rm Var}[\hat{N}_{\rm I}^{\rm out}(\pi)]}\right)^2}
\end{equation}
is easily shown to be 
\begin{equation}
\label{eqSNRQI}
{\rm SNR}_{\rm QI}  =  \frac{16\kappa_{\rm I}\kappa_{\rm S}\kappa_{\rm extra}\eta_{\rm D}MN_{\rm S}}{N_{\rm B}+N_{\rm el}},
\end{equation}
where $\eta_{\rm D}$ is the detector quantum efficiency, $N_{\rm el}$ is due to post-detection electronics noise, $\kappa_{\rm extra} < 1$ models all system nonidealities not otherwise accounted for, and $N_{\rm S} \ll 1$ plus $(G-1)N_{\rm B} \ll 1$ have been assumed.  

The preceding OPA receiver is not the optimum quantum measurement for QI target detection---its SNR is 3\,dB inferior to that of the optimum receiver \cite{guha09}---but no receiver structure is known that realizes that optimum quantum measurement.  Even so, the nonclassical PSCC of the SPDC's outputs makes ${\rm SNR}_{\rm QI}$ from Eq.~(\ref{eqSNRQI}) a factor of $1/N_{\rm S} \gg 1$ higher than what would result were the same receiver employed with a transmitter whose signal-idler mode pairs had a PSCC at the classical limit.  More importantly, the ideal ($\eta_{\rm D} = \kappa_{\rm I} = \kappa_{\rm extra} = 1$, $N_{\rm el} = 0$) OPA receiver's ${\rm SNR}_{\rm QI}$ is 3\,dB better than that of the optimum CI system of the same $MN_{\rm S}$ probe energy, 
\begin{equation}
\label{eqSNRCI}
{\rm SNR}_{\rm CI} = \frac{8\kappa_{\rm S}MN_{\rm S}}{N_{\rm B}},
\end{equation} 
under similar ideal conditions.

\begin{figure*}[htbp]
\centering
\subfigure{\includegraphics[width=4.2in]{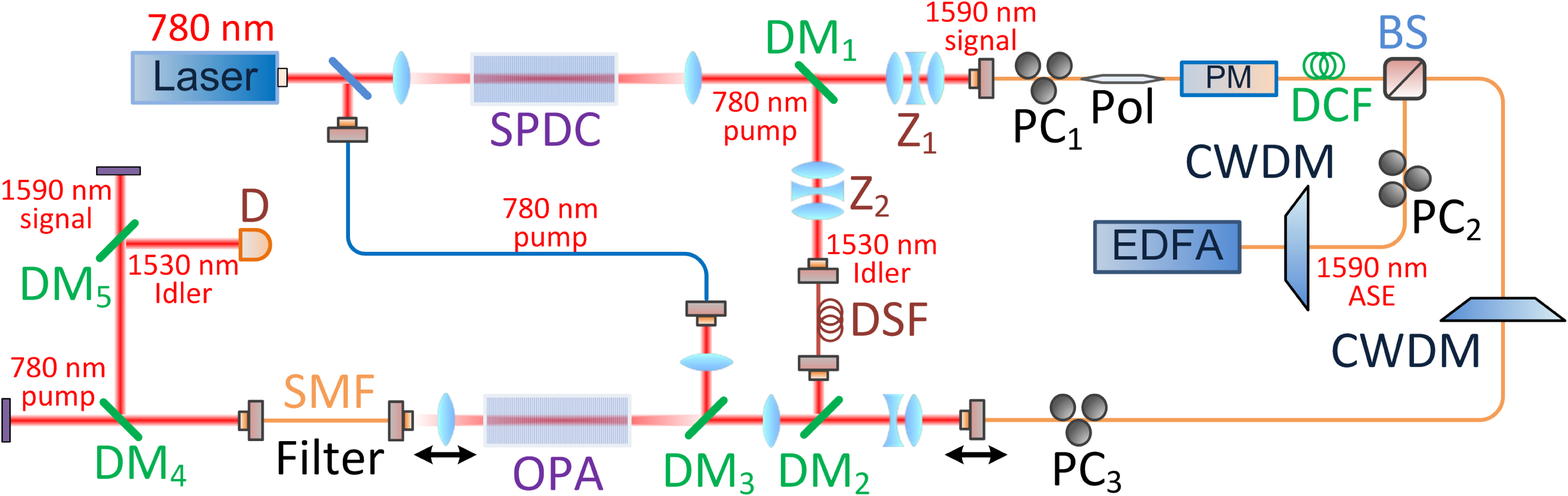}}
\subfigure{\includegraphics[width=2.8in]{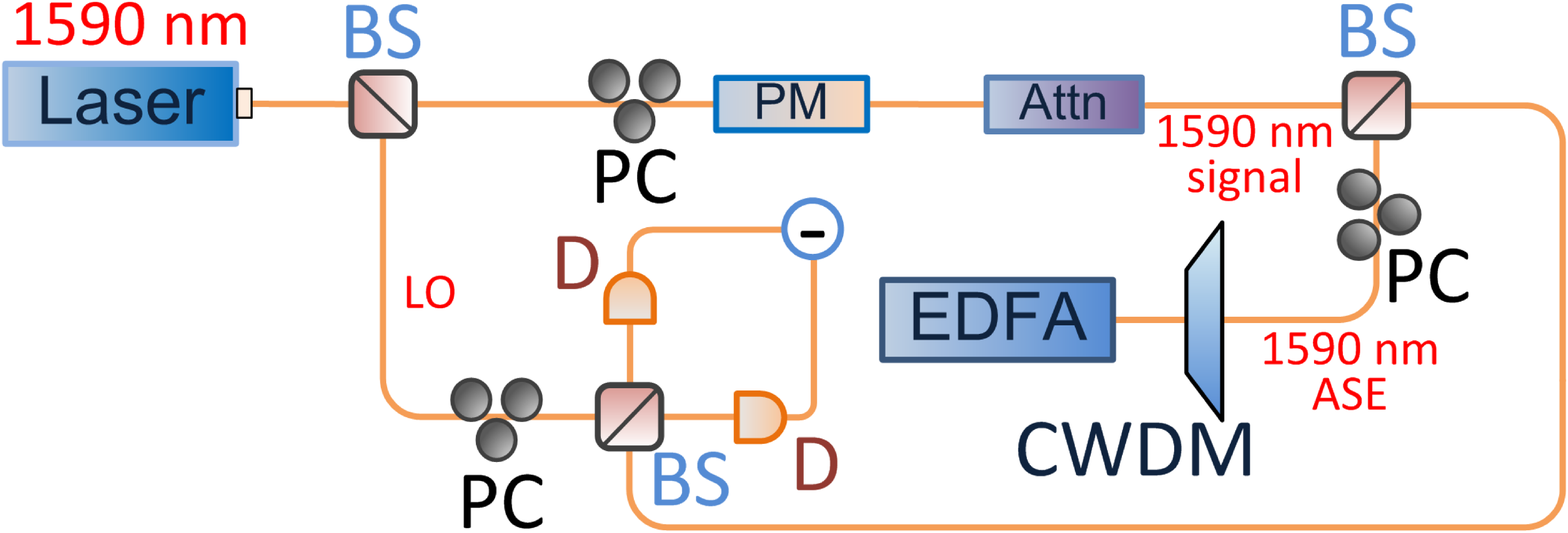}}
\caption{\label{figExp} (Color online) Experimental setups for quantum illumination (left) and classical illumination (right). DM: dichroic mirror; PC: polarization controller; Z: zoom lens; BS: beam splitter; SMF: single-mode fiber; SPDC: spontaneous parametric down-conversion; OPA: optical parametric amplifier; DSF: dispersion-shifted LEAF fiber; DCF: dispersion-compensating fiber; POL: polarizer; PM: phase modulator; EDFA: erbium-doped fiber amplifier; ASE: amplified spontaneous emission; CWDM: coarse wavelength-division multiplexer; D: detector; LO: local oscillator; Attn: attenuator.  Thin lines are optical fiber; thick lines are unguided propagation.}
\end{figure*}

Figure~\ref{figExp} (left) shows our experimental setup for QI. The SPDC source is a 4-cm-long type-0 magnesium-oxide doped  periodically-poled lithium niobate (MgO:PPLN) bulk crystal that is pumped by a 780\,nm diode laser which is followed by a tapered amplifier.  The pump is loosely focused at the PPLN crystal to guarantee a high signal-heralding efficiency \cite{bennink10}.  
The broadband entangled signal (1590\,nm center wavelength) and idler (1530\,nm center wavelength) beams are separated using a dichroic mirror (DM) that is highly reflective at the idler wavelength. The signal and idler beams pass through two zoom-lens systems, to adjust their confocal parameters for coupling into single-mode fiber (SMF) with maximum signal-heralding efficiency \cite{dixon14}. Ideally, every emitted signal photon would then herald an idler-photon companion, so that no probe photon is wasted. The idler is stored in a $\sim$30-meter-long low-dispersion, negligible-loss, LEAF-fiber spool whose ends are anti-reflection (AR) coated. A phase modulator imposes a 500\,Hz binary phase-shift keying (BPSK) modulation on the signal. A $\sim$4-meter-long dispersion-compensating fiber (DCF) overcompensates signal-beam dispersion in the SMF \cite{footnote1}.  The dispersion-overcompensated signal is combined with broadband amplified spontaneous emission (ASE) noise at a 50--50 beam splitter. The resulting noisy signal is filtered using a 16-nm-wide coarse wavelength-division multiplexer channel centered at 1590\,nm. The filtered signal is coupled into free space and fine-tuned in time delay using a prism (not shown). A convex-concave lens pair is employed to fine tune the signal's beam waist and location at the OPA. The signal beam is then combined with the retained idler on a DM that is highly reflective at the idler wavelength. Half-wave and quarter-wave plates (not shown) are used to adjust idler's polarization before combining. The combined  signal and idler are united with the retained pump---whose pre-combining polarization is adjusted by half-wave and quarter-wave plates (not shown)---on a DM that is highly reflective above 1400\,nm.  Signal, idler, and pump are then injected into a second 4-cm-long type-0 MgO:PPLN bulk crystal that serves as the receiver's OPA. A spatial filter, consisting of two collimators and AR-coated single-mode fibers, is employed at the OPA's output to reject non-collinear modes. By means of a movable lens, $\sim$90\% throughput for the idler is realized. The spatially-filtered light is then guided to a pair of DMs that are highly reflective above 1400\,nm, to reject the pump, and a second pair of DMs that are highly reflective at the idler wavelength, to reject the signal, before the idler is focused into a free-space AR-coated InGaAs PIN detector with $84\%$ quantum efficiency. The detected idler power is of the order of 1\,nW. The resulting weak photocurrent is amplified using a low-noise transimpedance amplifier with $5\times10^9$\,V/A transimpedance gain, 1\,kHz bandwidth, and 6\,fA/$\sqrt{\rm Hz}$ current-noise spectrum at its input. The amplifier's output voltage is directed to either an oscilloscope or a low-noise 100 kHz fast Fourier transform spectrum analyzer (FFT-SA) for SNR measurements. 

Figure~\ref{figExp} (right) shows our experimental setup for CI.  Light from a 1590\,nm laser is guided to a 99--1 beam splitter, whose 1\% output serves as the signal while its 99\% output supplies the local oscillator for homodyne reception. The signal's strength is further reduced by a tunable digital attenuator, after which 70\,kHz BPSK modulation is applied. The modulated signal is mixed on a 50--50 beam splitter with broadband ASE noise, as done in the QI experiment. The noisy signal is combined with the local oscillator on a 50--50 beam splitter, whose two output arms are detected by a balanced receiver with $ \sim $80\% quantum efficiency, an internal transimpedance gain of $2 \times 10^5$\,V/A, 120 kHz\,bandwidth, and negligible electrical noise (i.e., measurement is shot-noise limited). The voltage signal from the balanced receiver is directed to either the oscilloscope or the FFT-SA for SNR measurements.  

The presence of a target yields a sharp signal peak at the BPSK frequency on the FFT-SA, whereas a flat noise spectrum will be observed in the absence of a target, which we accomplish by manually blocking the signal before it mixes with the ASE noise. The spectral-peak amplitude (SPA) fluctuates owing to thermal and mechanical instabilities that cause random phase drifts between the QI receiver's signal and idler and between the CI receiver's signal and local oscillator. We are interested, however, in the maximum SPAs for QI and CI, as functions of $N_{\rm S}$ and $\kappa_{\rm S}$. These are obtained as follows.

To measure the maximum QI SPA, we first optimize the signal-heralding efficiency, spatial-filter collection efficiency, polarizations of all beams, and signal-idler relative delay, and then set the source brightness, channel transmissivity, and OPA gain to desired values. We use the FFT-SA to record the maximum SPA over 32,767 samples and repeat the measurement five times before switching to different experimental settings. During this $\sim$10-minute process, peak-to-peak signal-power fluctuations before the phase modulator was found to be $\pm$3\%. 

To measure the maximum CI SPA, we first optimize the local oscillator's polarization and then employ the same measurement procedure used for QI. In the absence of a target, we block the signal and directly measure the QI and CI noise spectral densities at their respective BPSK frequencies. The resolution bandwidth is 977\,mHz for all measurements. 

To connect the maximum-SPA and the noise spectral-density measurements for QI and CI with Eqs.~(\ref{eqSNRQI}) and (\ref{eqSNRCI}), we need to convert each maximum SPA to its corresponding intensity-modulation amplitude (IMA), defined by the numerators of those two equations. To do so we use an oscilloscope to record IMA histograms after performing the parameter optimizations described in the preceding paragraphs. CI's IMA histograms are measured using strong signal light, whereas QI's IMA histograms are taken at the same signal power levels used in its maximum-SPA measurements. The necessity of employing strong signal light in obtaining  CI IMA histograms is due to the CI homodyne receiver's current amplifier having a much lower gain than that of its QI-setup counterpart. Each IMA histogram consists of 10,000 samples and is captured within $\sim$10 minutes. We repeated the histogram measurements three times each for QI and CI, and found that the recorded maximum IMA only fluctuates $\pm$1.8\%, which is consistent with the signal power fluctuations. The maximum IMA in the histogram is proportional to the maximum SPA, enabling us to convert each SPA into its corresponding IMA, calculate the SNR, and compare with theory. 

\begin{figure}
\includegraphics[width=2.8in]{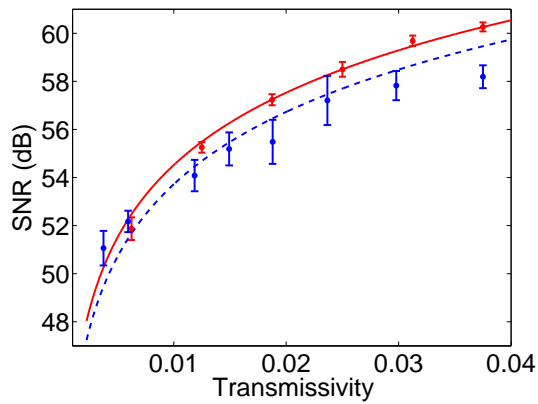}
\caption{\label{FigSNR_vs_kappa_S} (Color online) SNR versus transmissivity measurements for QI (top) and CI (bottom). In QI, $ G - 1 = 7.4 \times 10^{-5}$ and $N_{\rm S} =  3\times 10^{-4}$. CI has a photon flux of $ 5.67\times 10^8 $/s. $N_{\rm B} = 95$ at both receivers. Curves: QI theory (solid), CI theory (dashed). Error bars represent $\pm$1 measurement standard deviation and include the effect of $\pm$3\% signal-power fluctuations.} 
\end{figure}

Figure~\ref{FigSNR_vs_kappa_S} shows the measured QI and CI SNRs as functions of their channel transmissivities from the outputs of the transmitters to the inputs of the receivers. Each data point represents five consecutive SNR measurements, and the error bars denote $\pm$1 measurement standard deviation with signal-power fluctuations accounted for.  A polarization controller (PC$_1$) followed by a polarizer was used to vary the channel transmissivity in the QI experiment, whereas  a tunable attenuator was used to vary the channel transmissivity in the CI experiment. The SPDC source had brightness $N_{\rm S} =  3\times 10^{-4}$ and $W= 1.89$\,THz phase-matching bandwidth for all QI measurements, while all CI measurements were taken with the equivalent photon flux of $WN_{\rm S} = 5.67\times 10^8$\,photons/s. In both experiments, we had $N_{\rm B} =  95$ at the receivers, which is 69\,dB stronger than the returned signal power when the target was present and $\kappa_{\rm S} = 0.038$. The QI receiver had an estimated $\kappa_{\rm extra}\sim$0.8, representing the combined effects of the measured $\sim$90\% spatial-filter collection efficiency and $\sim$10\% additional nonidealities arising from imperfect signal-heralding efficiency and dispersion compensation, confirmed separately in heralding efficiency measurements using InGaAs avalanche photodiodes. 

The SNR measurements in Figure~\ref{FigSNR_vs_kappa_S} are in good agreement with theory. Remarkably, QI's SNR exceeds that of the optimum CI scheme by 20\% at $3.8$\% channel transmissivity, where the background light is 69\,dB stronger than the target return.  Also, the QI system continues to offer a performance advantage even when it suffers a 19\,dB transmission loss.

\begin{figure}[h]
\includegraphics[width=2.8in]{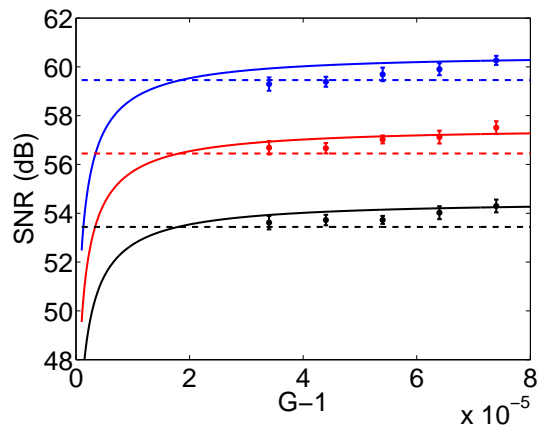}
\caption{\label{FigSNR_vs_Gain} (Color online) SNR versus OPA gain measurements for QI. $ N_{\rm S} =  3\times 10^{-4}$ (top), $ N_{\rm S} =  1.5\times 10^{-4}$ (middle), and $ N_{\rm S} =  7.5\times 10^{-5}$ (bottom). $ N_{\rm B} = 95$ at the receivers. Solid curves:  SNR theory for QI using $\kappa_{\rm extra} = 0.8$.  Dashed lines: SNRs for theoretically-optimum CI systems at photon flux $WN_{\rm S}$ with $W = 1.89\,$THz.} 
\end{figure}

QI's SNR is known to be a function of its OPA gain \cite{guha09}. A sufficiently high OPA gain guarantees that the idler's optical noise overwhelms post-detection electronics noise, but a high OPA gain violates the $(G-1)N_{\rm B}\ll 1 $ assumption and thus introduces additional thermal-state noise \cite{footnote2}.  Figure~\ref{FigSNR_vs_Gain} plots QI's measured SNR versus OPA gain for three different source brightnesses.  All measurements were performed with 14\,dB channel loss and $N_{\rm B} = 95$ at the receiver.   At $ N_{\rm S} =   7.5\times 10^{-5}$, this background is 75\,dB stronger than the target return. For all three $N_{\rm S}$ values, QI provided a 20\% SNR improvement over the theoretically-optimum CI performance when the QI receiver's OPA gain satisfied $G - 1 = 7.4 \times 10^{-5}$. The solid curves are theoretical results for QI's SNR obtained using $\kappa_{\rm extra} = 0.8$. The departure from the theoretical predictions at low OPA gains results from slow degradations of the signal-heralding efficiency that are due to mechanical instabilities, and signal-idler delay mismatch caused by thermal drifts. These instabilities occur on time scales of minutes, and have been confirmed separately in signal-heralding efficiency measurements and fine tuning of the prism to recover the maximum QI SNR.

Our QI experiment's SNR advantage over CI has been reaped even though the returned signal and the retained idler are in a classical state, i.e., the channel has broken the initial signal-idler entanglement.  That initial entanglement will be broken when loss and noise are such that the PSSCs at the QI receiver's OPA input satisfy $|\langle\hat{a}_{\rm S}^{\rm in}(\varphi)\hat{a}_{\rm I}^{\rm in} \rangle |^2 \le  \langle\hat{a}_{\rm S}^{{\rm in}\dagger}(\varphi)\hat{a}_{\rm S}^{\rm in}(\varphi)\rangle \langle\hat{a}_{\rm I}^{{\rm in}\dagger}\hat{a}_{\rm I}^{\rm in}\rangle$.  In our experiment, $|\langle\hat{a}_{\rm S}^{\rm in}(\varphi)\hat{a}_{\rm I}^{\rm in} \rangle |^2$ is 34--41\,dB below that classical-state upper limit. This result proves that a quantum resource can indeed yield an appreciable performance improvement in sensing, even though quantum decoherence destroys the initial nonclassicality, and thus it may open a window for applying QIP in practical (lossy, noisy) situations. 

This research was supported by ARO grant number W911NF-10-1-0430 and ONR grant number N00014-13-1-0774.


\begin{thebibliography}{}
\bibitem{footnote1}Overcompensation is done so that when the signal and idler are combined at the OPA  their respective photons have both been stretched from the 0.2\,ps durations they had at the SPDC source to $\sim$3\,ps, with one beam having normal dispersion and the other anomalous dispersion \cite{zhang13} so that the cross correlation of the signal-idler Gaussian state is recovered \cite{shapiro10}.
\bibitem{footnote2}As a practical matter, one should limit the OPA gain to avoid saturating the QI receiver's current amplifier.
\end{thebibliography}
\end{document}